\documentstyle[12pt]{article}
\begin{document}
\title{The transition of a gravitationally radiating,  dissipative fluid, to equilibrium}
\author{L. Herrera$^{1}$\thanks{e-mail: lherrera@usal.es},
 A. Di Prisco$^{2}$\thanks{e-mail: alicia.diprisco@ciens.ucv.ve}, and J. Ospino$^{1,3}$\thanks{email: j.ospino@usal.es}
\\\
\small{$^1$ Instituto Universitario de Fisica Fundamental y Matematicas}\\
\small{Universidad de Salamanca, Salamanca, Spain}\\
\small{$^2$Escuela de F\'{\i}sica, Facultad de Ciencias,} \\
\small{Universidad Central de Venezuela, Caracas, Venezuela.}\\
\small{$^3$ Departamento de Matem\'atica Aplicada, Universidad de Salamanca, Salamanca, Spain.}
\\
}
\maketitle

\vspace{-0.5cm}
\begin{abstract}
We describe the transition of a gravitationally radiating,  axially and reflection symmetric dissipative fluid, to a non--radiating state. It is shown that very shortly after the end of the radiating regime, at a time scale of  the order the thermal relaxation time, the thermal adjustment time and the hydrostatic time (whichever is larger),  the system reaches the equilibrium state.  This result is at variance with all the studies carried out in the past, on gravitational radiation  outside the source,  which strongly suggest that  after a radiating period, the conditions  for a return to a static case, look rather forbidding. As we shall see, the reason for such a discrepancy resides in the fact that  some elementary, but essential, physical properties of the source, have been overlooked in these latter studies.
\end{abstract}

\maketitle

\section{Introduction}
Very powerful methods to study gravitational radiation, beyond the well known linear approximation, were put forward in the sixties \cite{7, 8, NP, BR}. Besides the many fundamental results obtained from these methods, their main merit consists in including non--linear effects, which are known to play a very important role in general relativity.  All these approaches describe the gravitational radiation outside its source, generally very far from it, to avoid the appearance of caustics and similar pathologies.

Among the wealth of relevant results obtained by means of  the methods mentioned above, there is one which has attracted the attention of researchers for many decades, namely:  the existence of gravitational wave tails, which implies that the Huygens's principle does not apply to gravitational  waves, if non--linearities are taken into account. The bibliography on this issue is overly lengthy huge, and just as a very restricted and incomplete sample, we shall refer the reader to \cite{DB, KN, BoNP, NPn,  CTJN, Bo, BS, MBF}.

One inmediate consequence of  the violation of the   Huygens's principle, is the fact that once the system stops to radiate, it does not return back to the static regime, but instead, enters into what Bondi calls ``time dependent systems without news'', i.e. there is no gravitational radiation, but the spacetime variables depend on time.

In a recent work \cite{HDOC}, we have studied the transition from a static situation to a non--equilibrium (radiative) one. 
In this work we shall study the inverse problem, i.e. the possible transition from a radiative to a static regime. However,  unlike the references mentioned above,  we shall focus on the source of the gravitational radiation, instead of the asymptotic structure of spacetime.

Gravitational wave tails are supposed to come from scattering of outgoing waves off the spacetime, this could lead to the conclusion that the late-time scattered gravitational waves will presumably invalidate the assumption of perfect
equilibrium, once the radiation process has ceased.  However, as we shall see here, this is not so. Indeed, the above conclusion  is true {\bf only} if  the physical processes within the source are neglected.
Equilibrium implies static, if one takes into consideration the source, this is the main point of our research.

Thus, we shall consider an evolving system, consisting  of a fluid distribution, which due to the changes of its  multipole moments, produces gravitational radiation. We shall next assume that it  ceases to radiate at some time (say $t=0$). For this to happen it is necessary that at $t=0$, any fluid element reaches the equilibrium, implying that the hydrostatic equilibrium equations (Eqs.(21,22) in \cite{static}) are satisfied. Shortly after  $t=0$, at a time scale of the order of, or larger than the thermal relaxation time, the thermal adjustment time and the hydrostatic time \cite{astr1},\cite{astr2}, all transient effects have vanished. It is from this time and on, that we shall evaluate our system, to check if there is any potential impediment, for the  transition from a radiative regime  to the equilibrium situation, to occur (putting aside the short time required for the transient effects to  vanish).
 
As our main result, we obtain that such a transition is indeed  possible, after small time scale of the order of the thermal relaxation time, the thermal adjustment time and the hydrostatic time. Obviously, a static source emerging after the end of the radiation period,  would produce a static exterior spacetime, which is in contradiction with the result mentioned above, about the  non validity of the   Huygens's principle. As we shall discuss in the last section, such discrepancy is related to the fact that the frictional forces existing  within the source, and responsible for the rapid decay of transient effects, are not taken into account in the studies focusing in the spacetime far  from the sources.

In our study  we shall heavily rely on a  general formalism  developed in \cite{1} using a framework based on the $1+3$ approach \cite{21cil, n1, 22cil, nin}. Accordingly,  in order to avoid the rewriting of some of the equations, we shall frequently refer to \cite{1} and \cite{HDOC}, however  we warn  the reader of some important changes in the notation with respect to the first of these references.
\section{Basic definitions and notation}
In this section we shall deploy, without giving details,  all the variables required for our study, some details of the calculations are given  in \cite{1} and \cite{HDOC}, and therefore we shall omit them here.
\subsection{The metric, the source, and the kinematical variables}

We shall consider,  axially (and reflection) symmetric sources. For such a system the  line element may be written in ``Weyl spherical coordinates'' as:

\begin{equation}
ds^2=-A^2 dt^2 + B^2 \left(dr^2
+r^2d\theta^2\right)+C^2d\phi^2+2Gd\theta dt, \label{1b}
\end{equation}
where $A, B, C, G$ are positive functions of $t$, $r$ and $\theta$. We number the coordinates $x^0=t, x^1=r, x^2= \theta, x^3=\phi$.

We shall assume that  our source is filled with an anisotropic and dissipative fluid. We are concerned with either bounded or unbounded configurations. In the former case we should further assume that the fluid is bounded by a timelike surface $S$, and junction (Darmois) conditions should be imposed there.

The energy momentum tensor may be written in the ``canonical'' form, as 
\begin{equation}
{T}_{\alpha\beta}= (\mu+P) V_\alpha V_\beta+P g _{\alpha \beta} +\Pi_{\alpha \beta}+q_\alpha V_\beta+q_\beta V_\alpha.
\label{6bis}
\end{equation}

The above is the canonical, algebraic decomposition of a second order symmetric tensor with respect to unit timelike vector, which has the standard physical meaning when $T_{\alpha \beta}$ is the energy-momentum tensor describing some energy distribution, and $V^\mu$ the four-velocity assigned by certain observer.

With the above definitions it is clear that $\mu$ is the energy
density (the eigenvalue of $T_{\alpha\beta}$ for eigenvector $V^\alpha$), $q_\alpha$ is the  heat flux, whereas  $P$ is the isotropic pressure, and $\Pi_{\alpha \beta}$ is the anisotropic tensor. We are considering an Eckart frame  where fluid elements are at rest.

Since we choose the fluid to be comoving in our coordinates, then
\begin{equation}
V^\alpha =\left(\frac{1}{A},0,0,0\right); \quad  V_\alpha=\left(-A,0,\frac{G}{A},0\right).
\label{m1}
\end{equation}

We shall next define a canonical  orthonormal tetrad (say  $e^{(a)}_\alpha$), by adding to the four--velocity vector $e^{(0)}_\alpha=V_\alpha$, three spacelike unitary vectors (these are denoted by  {$\bf K$}, {$\bf L$}, {$\bf S$} respectively,  in \cite{1})

\begin{equation}
e^{(1)}_\alpha=(0,B,0,0); \quad  e^{(2)}_\alpha=\left(0,0,\frac{\sqrt{A^2B^2r^2+G^2}}{A},0\right),
\label{7}
\end{equation}

\begin{equation}
 e^{(3)}_\alpha=(0,0,0,C),
\label{3nb}
\end{equation}
with $a=0,\,1,\,2,\,3$ (latin indices labeling different vectors of the tetrad)

The  dual vector tetrad $e_{(a)}^\alpha$  is easily computed from the condition 
\begin{equation}
 \eta_{(a)(b)}= g_{\alpha\beta} e_{(a)}^\alpha e_{(b)}^\beta, \qquad e^\alpha_{(a)}e_\alpha^{(b)}=\delta^{(b)}_{(a)},
\end{equation}
where $\eta_{(a)(b)}$ denotes the Minkowski metric.

In the above, the tetrad vector $e_{(3)}^\alpha=(1/C)\delta^\alpha_\phi$ is parallel to
the only admitted Killing vector (it is the unit tangent to the orbits of the
group of 1--dimensional rotations that defines axial symmetry). The other two
basis vectors $e_{(1)}^\alpha,\,e_{(2)}^\alpha$ define the two {\it unique}
directions that are orthogonal to the 4--velocity and to the Killing vector.

It can be shown that  the anisotropic tensor  may be  expressed through three scalar functions defined as (see \cite{HDOC}):

\begin{eqnarray}
 \Pi _{(2)(1)}=e^\alpha_{(2)}e^\beta_{(1)} T_{\alpha \beta} 
, \quad , \label{7P}
\end{eqnarray}

\begin{equation}
\Pi_{(1)(1)}=\frac{1}{3}\left(2e^{\alpha}_{(1)} e^{\beta}_{(1)} -e^{\alpha}_{(2)} e^{\beta}_{(2)}-e^{\alpha}_{(3)} e^{\beta}_{(3)}\right) T_{\alpha \beta},
\label{2n}
\end{equation}
\begin{equation}
\Pi_{(2)(2)}=\frac{1}{3}\left(2e^{\alpha}_{(2)} e^{\beta}_{(2)} -e^{\alpha}_{(3)} e^{\beta}_{(3)}-e^{\alpha}_{(1)} e^{\beta}_{(1)}\right) T_{\alpha \beta}.
\label{2nbis}
\end{equation}

We may write the heat flux vector in terms of  the two tetrad components $q_{(1)}$ and $q_{(2)}$:
\begin{equation}
q_\mu=q_{(1)}e_{\mu}^{(1)}+q_{(2)}e_{\mu}^{(2)}
\label{qn1}
\end{equation}
or, in coordinate components (see \cite{1})
\begin{equation}
q^\mu=\left(\frac{q_{(2)} G}{A \sqrt{A^2B^2r^2+G^2}},  \frac{q_{(1)}}{B}, \frac{Aq_{(2)}}{\sqrt{A^2B^2r^2+G^2}}, 0\right)
,\label{q}
\end{equation}
\begin{equation}
 q_\mu=\left(0, B q_{(1)}, \frac{\sqrt{A^2B^2r^2+G^2}q_{(2)}}{A}, 0\right).
\label{qn}
\end{equation}

Of course, all the above quantities depend,  in general, on $t, r, \theta$.

For the kinematical variables we have the following expressions  (see  \cite{1, HDOC}).

For the four acceleration we have
\begin{equation}
a_\alpha=V^\beta V_{\alpha;\beta}=a_{(1)}e_{\mu}^{(1)}+a_{(2)}e_{\mu}^{(2)},
\label{a1n}
\end{equation}
with
\begin{equation}
a_{(1)}= \frac {A^\prime }{AB };\quad a_{(2)}=\frac{A}{\sqrt{A^2B^2r^2+G^2}}\left[\frac {A_{,\theta}}{A}+\frac {G}{A^2}\left(\frac{\dot G}{G}-\frac{\dot A}{A}\right)\right],
\label{accn}
\end{equation}
 where the dot  and the prime denote derivatives with respect to $t$ and $r$ respectively. 

For the expansion scalar
\begin{eqnarray}
\Theta&=&V^\alpha_{;\alpha}=\frac{1}{A}\left(\frac{2 \dot B}{B}+\frac{\dot C}{C}\right) \nonumber\\
&+&\frac{G^2}{A\left(A^2 B^2 r^2 + G^2\right)}\left(-\frac{\dot A}{A}-\frac{\dot B}{B}+\frac{\dot G}{G}\right).
\label{thetan}
\end{eqnarray}

Next, the shear tensor
\begin{equation}
\sigma_{\alpha \beta}=\sigma_{(a)(b)}e^{(a)}_\alpha e^{(b)}_\beta=V_{(\alpha;\beta)}+a_{(\alpha}
V_{\beta)}-\frac{1}{3}\Theta h_{\alpha \beta}, \label{acc}
\end{equation}
may be  defined through two independent tetrad components (scalars)  $\sigma_{(1)(1)}$ and $\sigma_{(2)(2)}$, 
which may be written in terms of the metric functions and their derivatives as (see \cite{HDOC, 1}):
\begin{eqnarray}
\sigma_{(1) (1)}&=&\frac{1}{3A}\left(\frac{\dot B}{B}-\frac{\dot C}{C}\right)\nonumber \\
&+&\frac{G^2}{3A\left(A^2 B^2 r^2 + G^2\right)}\left(\frac{\dot A}{A}+\frac{\dot B}{B}-\frac{\dot G}{G}\right),
 \label{sigmasI}
\end{eqnarray}
\begin{eqnarray}
\sigma_{(2)(2)}&=&\frac{1}{3A}\left(\frac{\dot B}{B}-\frac{\dot C}{C}\right)\nonumber \\ &+&\frac{2G^2}{3A\left (A^2 B^2 r^2 + G^2\right)}\left(-\frac{\dot A}{A}-\frac{\dot B}{B}+\frac{\dot G}{G}\right)
\label{sigmas}.
\end{eqnarray}

Finally,  for the vorticity tensor 
\begin{equation}
\Omega
_{\beta\mu}=\Omega_{(a)(b)}e^{(a)}_\beta
e^{(b)}_\mu,
\end{equation}
 we find that it is determined by a single basis component:
\begin{equation}
\Omega_{(1)(2)} = -\Omega_{(2)(1)}=-\Omega,
\label{omegan}
\end{equation}

where the scalar function $\Omega$ is given by
\begin{equation}
\Omega =\frac{G(\frac{G^\prime}{G}-\frac{2A^\prime}{A})}{2B\sqrt{A^2B^2r^2+G^2}}.
\label{no}
\end{equation}

Now, from the regularity conditions, necessary to ensure elementary flatness in the vicinity of  the axis of symmetry, and in particular at the center (see \cite{1n}, \cite{2n}, \cite{3n}), we should require  that as $r\approx 0$
\begin{equation}
\Omega=\sum_{n \geq1}\Omega^{(n)}(t,\theta) r^{n},
\label{sum1}
\end{equation}
implying, because of (\ref{no}) that in the neighborhood of the center
\begin{equation}
G=\sum_{n\geq 3}G^{(n)}(t, \theta) r^{n}.
\label{sum1}
\end{equation}

\subsection{The electric and magnetic part of the Weyl tensor and the super--Poynting vector}
Let us now introduce the electric ($E_{\alpha\beta}$) and magnetic ($H_{\alpha\beta}$) parts of the Weyl tensor ( $C_{\alpha \beta
\gamma\delta}$),  defined as usual by
\begin{eqnarray}
E_{\alpha \beta}&=&C_{\alpha\nu\beta\delta}V^\nu V^\delta,\nonumber\\
H_{\alpha\beta}&=&\frac{1}{2}\eta_{\alpha \nu \epsilon
\rho}C^{\quad \epsilon\rho}_{\beta \delta}V^\nu
V^\delta\,.\label{EH}
\end{eqnarray}

The electric part of the Weyl tensor has only three independent non-vanishing components, whereas only two components define the magnetic part. Thus  we may  write these two  tensors, in terms of five  tetrad components  (${\mathcal{E}}_{(1)(1)}, {\mathcal{E}}_{(2)(2)}$, ${\mathcal{E}}_{(1)(2)}, H_{(1)(3)}, H_{(3)(2)}$), respectively as (see \cite{1} for details):

\begin{eqnarray}
E_{\alpha\beta}&=&\left[\left(2{\mathcal{E}}_{(1)(1)}+{\mathcal{E}}_{(2)(2)}\right) \left(e^{(1)}_\alpha
e^{(1)}_\beta-\frac{1}{3}h_{\alpha \beta}\right)\right] +\left[\left(2{\mathcal{E}}_{(2)(2)}+{\mathcal{E}}_{(1)(1)}\right)\left (e^{(2)}_\alpha
e^{(2)}_\beta-\frac{1}{3}h_{\alpha \beta}\right)\right]\nonumber \\&+&{\mathcal{E}}_{(2)(1)} \left(e^{(1)}_\alpha
e^{(2)}_\beta+e^{(1)}_\beta
e^{(2)}_\alpha\right), \label{E'}
\end{eqnarray}

\noindent

and
\begin{eqnarray}
H_{\alpha\beta}=H_{(1)(3)}\left(e^{(1)}_\beta
e^{(3)}_\alpha+e^{(1)}_\alpha
e^{(3)}_\beta \right)\nonumber \\+H_{(2)(3)}\left(e^{(3)}_\alpha
e^{(2)}_\beta+e^{(2)}_\alpha
e^{(3)}_\beta \right)\label{H'}.
\end{eqnarray}

Next, in the well known  irreducible decomposition of the Riemann tensor, in terms of the Weyl tensor, the Ricci tensor and the curvature scalar, we shall replace these two latter geometrical quantities, by their expressions in terms of the energy momentum tensor, as implied by the Einstein equations. Doing so, the obtained expression for the Riemann  tensor embodies the Einstein equations (see \cite{HDOC} for details).

Also, from  the Riemann tensor we may define  three tensors $Y_{\alpha\beta}$, $X_{\alpha\beta}$ and
$Z_{\alpha\beta}$ as

\begin{equation}
Y_{\alpha \beta}=R_{\alpha \nu \beta \delta}V^\nu V^\delta,
\label{Y}
\end{equation}
\begin{equation}
X_{\alpha \beta}=\frac{1}{2}\eta_{\alpha\nu}^{\quad \epsilon
\rho}R^\star_{\epsilon \rho \beta \delta}V^\nu V^\delta,\label{X}
\end{equation}
and
\begin{equation}
Z_{\alpha\beta}=\frac{1}{2}\epsilon_{\alpha \epsilon \rho}R^{\quad
\epsilon\rho}_{ \delta \beta} V^\delta,\label{Z}
\end{equation}
 where $R^\star _{\alpha \beta \nu
\delta}=\frac{1}{2}\eta_{\epsilon\rho\nu\delta}R_{\alpha
\beta}^{\quad \epsilon \rho}$  and $\epsilon _{\alpha \beta \rho}=\eta_{\nu
\alpha \beta \rho}V^\nu$.

The above tensors in turn, may be  decomposed, so that each of them is described through  four scalar functions known as structure scalars \cite{sc}. These are (see \cite{1} for details)
\begin{eqnarray}
Y_T&=&4\pi(\mu+3P), \qquad X_T=8\pi \mu, \label{ortc1}\\
Y_I&=&3{\mathcal{E}}_{(1)(1)}-12\pi \Pi_{(1)(1)},\quad X_I=-3{\mathcal{E}}_{(1)(1)}-12\pi \Pi_{(1)(1)}
\nonumber\\
Y_{II}&=&3{\mathcal{E}}_{(2)(2)}-12\pi \Pi_{(2)(2)},\quad X_{II}=-3{\mathcal{E}}_{(2)(2)}-12\pi \Pi_{(2)(2)}, \nonumber\\
Y_{III}&=&{\mathcal{E}}_{(2)(1)}-4\pi \Pi_{(2)(1)}, \quad X_{III}=-{\mathcal{E}}_{(2)(1)}-4\pi \Pi_{(2)(1)}.\nonumber
\end{eqnarray}
and

\begin{eqnarray}
Z_I=(H_{(1)(3)}-4\pi q_{(2)});\quad Z_{II}=(H_{(1)(3)}+4\pi  q_{(2)}) \nonumber \\ Z_{III}=(H_{(2)(3)}-4\pi q_{(1)}); \quad  Z_{IV}=(H_{(2)(3)}+4\pi q_{(1)}). \label{Z2}
\end{eqnarray}

\subsection{The variables}
To summarize, the whole set of variables fully describing our system are: 
\begin{itemize}
\item The metric variables $A, B, C, G$ (or the tetrad vectors).
\item The kinematical variables $a_{(1), (2)}$, $\Theta$, $\sigma_{(1)(1)}$, $\sigma_{(2)(2)}$, $\Omega$.
\item  The physical variables describing the fluid distribution  $\mu, P, \Pi_{(1)(1)}, \Pi_{(2)(2)}, \Pi_{(2)(1)}$.
\item  The dissipative fluxes $q_{(1)}, q_{(2)}$.
\item  The three scalars defining the electric part of the Weyl tensor ${\mathcal{E}}_{(1)(1)}, {\mathcal{E}}_{(2)(2)}, {\mathcal{E}}_{(2)(1)}$ and the two scalars defining the magnetic part of the Weyl tensor $H_{(1)(3)}, H_{(2)(3)}$.
\end{itemize}

\section{The  equations}

To determine the evolution of our system, we shall need a set of differential equations, which in  the context of the $1+3$ approach, is a system of first order equations for the variables listed above.  Besides we shall need  a transport equation for the dissipative processes. Obviously, for any specific model we should also have available equations of state linking  different fluid variables. However, since our  study is not related to any specific model, we shall not need to refer to any specific equation of state.

In what follows, we shall point out  the origin of different set of equations, without specifying them further, since they are explicitly written in \cite{1}, from where we shall import them, as required.
\subsection{The transport equation}

In the presence of dissipative processes, and in order to avoid the drawbacks generated by the standard (Landau--Eckart) irreversible thermodynamics \cite{17}, \cite{67}, (see \cite{63}-\cite{66} and references therein)  we need a transport equation derived from  a causal  dissipative theory. In the past  (see \cite{HDOC, 1} and references therein)  we have  resorted to 
M\"{u}ller-Israel-Stewart second
order phenomenological theory for dissipative fluids \cite{18, 19, 20, 21}). However, in this work, the study of the system begins once the thermal equilibrium has been established, and therefore we have  no need of any specific transport equation.
\subsection{The differential equations for all the variables}

First, we have a first order differential equation system, relating the metric variables with the kinematical variables, these are the equations (\ref{accn},  \ref{thetan}, \ref{sigmasI}, \ref{sigmas}, \ref{no}), or in a condensed form
\begin{equation}
V_{\alpha;\beta}=\sigma_{\alpha \beta}+\Omega_{\alpha \beta}-a_\alpha V_\beta+\frac{1}{3}h_{\alpha \beta}\Theta.\label{ps}
\end{equation}

Next, the integrability conditions of (\ref{ps}) read
\begin{equation}
V_{\alpha ;\beta ; \nu}- V_{\alpha;\nu;\beta}=R^{\mu}_{\alpha
\beta \nu}V_\mu. 
\label{Ricci}
\end{equation}
These last equations provide evolution equations for  $\Theta$,  $\sigma_{\alpha \beta}$, and $\Omega$. Besides, they provide differential constraints relating dissipatives fluxes with derivatives of the kinematical variables, as well as contraints relating the magnetic part of the Weyl tensor with derivatives of the kinematical variables. These are the equations (B1)--(B9) in \cite{1}.

The integrability conditions of (\ref{Ricci}) are just the Bianchi identities, which provide evolution equations for $X_{I}, X_{II}, X_{III}, Y_{I}, Y_{II},Y_{III}, H_{(1)(3)}, H_{(2)(3)}$, as well as differential constraints for the spatial derivatives of the above quantities. These are the equations (B10)--(B18) in \cite{1}.

Finally, it could be useful to write the Bianchi identities as ``the conservation laws'' $T^\mu_{\nu; \mu}=0$. These are the equations (A6),(A7) in \cite{1}.

\section{Leaving the radiative regime}

From very simple physical considerations, it should be obvious that for the radiation regime to stop, the fluid distribution within the source must reach the hydrostatic and thermal equilibrium. Such a state, requires the fulfillment of hydrostatic equilibrium equations, as well as the fulfillment of Tolman conditions for thermal equilibrium \cite{Tolman}, and $ H_{(1)(3)}= H_{(2)(3)}=0$.

Then, shortly after the end of the radiation regime, where shortly means in a time of the order of (or larger than) the hydrostatic time, the thermal relaxation time and the thermal adjustment time, transient effects should have been faded away. 

At this point  it is worth recalling, that: 
\begin{itemize}
 \item The hydrostatic time is the typical time in which a fluid element reacts on a slight perturbation of hydrostatic equilibrium, it is basically of the order of magnitude of the time taken by a sound wave to propagate through the whole fluid distribution. 
\item The thermal relaxation time is the time taken by the system to return to the steady state in the heat flux (whether of thermodynamic equlibrium or not), after it has been removed from it.
\item Finally, the thermal adjustment time is the time it takes a fluid element to adjust thermally to its surroundings.  It is, essentially, of the order of magnitude of the time  required for a significant change in the temperature gradients.
\end{itemize}

From the above comments  it is clear that, once the radiation process has ended, and after a time period larger than the three time scales defined before, we have that :
\begin{itemize}
\item  The kinematical quantities $\Omega (G), \Theta, \sigma_{(1) (1)},\sigma_{(2) (2)}$  vanish, as well as the dissipative fluxes $q_{(1)}, q_{(2)}$. The vanishing of kinematical variables imply at once that first order time derivatives of the metric variables $ B, C$ vanish.

\item From the above conditions, it follows at once from (A6) in \cite{1}, that $\dot \mu=0$.
\item Since we want the equilibrium state to hold for a finite period of time after it was reached, we need to impose $\dot G=0$ (the vanishing of the ``source'' news function), which as shown in \cite{HDOC}, ensures that equilibrium is maintained. Otherwise, the system shall leave the equilibrium. 

\end{itemize}

Then, we have for the four acceleration
\begin{equation}
a_{(1)}= \frac {A^\prime }{AB };\quad a_{(2)}=\frac {A_{,\theta}}{ABr},
\label{accitem}
\end{equation}

and
\begin{equation}
\dot \Theta=\dot \sigma_{(1) (1)}=\dot \sigma_{(2) (2)} =\dot \sigma_{(2) (1)}=\dot H_{(1)(3)}=\dot H_{(2)(3)}=0.
\label{ds}
\end{equation}

All this implies in its turn (see Eqs.(56--60, 67) in \cite{HDOC}), that
\begin{equation}
\mu=\mu_{(eq)},\quad P=8\pi P_{(eq)},
\label{moeq}
\end{equation}

\begin{equation}
\Pi_{(1) (1)}=\Pi_{(1)(1) (eq)}, \quad \Pi_{(2)(2)}= \Pi_{(2)(2) (eq)},
\label{PiIoeq}
\end{equation}

\begin{equation}
 \Pi_{(2)(1)}=\Pi_{(2)(1)(eq)},
\label{PiKLoeq}
\end{equation}
where $eq$ stands for the value of the quantity at equilibrium.

Also, (see eqs.(68--71) in \cite{HDOC})

\begin{equation}
 {\cal E}_{(1)(1)}={\cal E}_{(1)(1)(eq)}, \quad {\cal E}_{(2)(2)}={\cal E}_{(2)(2)(eq)},
\label{EIoeq}
\end{equation}

\begin{equation}
{\cal E}_{(2)(1)}={\cal E}_{(2)(1) (eq)},
\label{EKLoeq}
\end{equation}
which impliy, because of (\ref{ortc1})
\begin{equation}
X_I=X_{I (eq)},\quad X_{II}=X_{II (eq)},\quad X_{III}=X_{III (eq)},
\label{xs}
\end{equation}
and
\begin{equation}
Y_I=Y_{I (eq)},\quad Y_{II}=Y_{II (eq)},\quad Y_{III}=Y_{III(eq)}.
\label{ys}
\end{equation}

In the above we have written down all the consequences emerging from the condition that the system ceases to radiate,  keeping this equilibrium state for at least a finite period. We shall next scrutinize all the equations listed in Section III, to check that the described transition, is not prohibited.

\section{Checking the compatibility of the return to the equilibrium state with the field equations}
Once the hydrostatic and thermal equilibrium of the source is assumed, in order to stop the generation of gravitational radiation, one should check if any of the equations described in Sec.III, forbids the transition to the static situation.

Let us start with the equations  (\ref{ps}), which, as has already been seen, implies the vanishing of the first time derivatives of the metric functions. 

Next we have the equations (\ref{Ricci}), which are the equations (B1--B9) in \cite{1}. They lead to the following constraints:
\begin{itemize}
\item (B1)$\rightarrow Y_T=a^\alpha_{;\alpha}$.
\item (B2)$\rightarrow-a^\delta
_{;\delta}+3(e^\mu_{(1)} e^\nu_{(1)} a_{\nu;\mu}+a_{(1)}^2)=Y_I$.
\item (B3)$\rightarrow a_{(1)} a_{(2)}+e_{(1)}^{(\mu}e_{(2)}^{\nu)}a_{\nu;\mu}=Y
_{III}$.
\item (B4)$\rightarrow -a^\delta
_{;\delta}+3(e^\mu_{(2)} e^\nu_{(2)} a_{\nu;\mu}+a_{(2)}^2)=Y_{II}$.
\item (B5)$\rightarrow e_{(1)}^{[\mu}e_{(2)}^{\nu]}a_{\mu;\nu}=0$.
\end{itemize}

 On the other hand, (B6--B7) relate the kinematical variables to the dissipative fluxes, while (B8, B9) relate those variables  to the two scalars that define the magnetic part of the Weyl tensor. Since, both,  the dissipative fluxes and the magnetic part of the Weyl tensor, as well as the kinematical variables, vanish, these four equations are identically satisfied.

Next we have the Bianchi identities, which are the equations (B10)--(B18) in \cite{1}), they imply:

\begin{itemize}
\item (B10)$\rightarrow \dot X_I=0$.
\item (B11)$\rightarrow \dot X_{III}=0$.
\item (B12)$\rightarrow \dot X_{II}=0$.
\item (B14)$\rightarrow 
-\frac{1}{3}X_{I,\beta}e^\beta_{(1)}-X_{III,\beta }e^\beta_{(2)}-\frac{1}{3}(2X_I+X_{II})(e^\beta_{(1) ;\beta}-a_\nu e^\nu_{(1)})-\frac{1}{3}(X_I+2X_{II})e^{(2)}_{\mu;\beta}e^\beta_{(2)} e^\mu_{(1)}
-X_{III}(e^{(2)}_{\mu;\beta}e^\mu_{(1)} e^\beta_{(1)}+e^\beta_{(2) ;\beta}-a_\beta e^\beta_{(2)})
=\frac{8\pi}{3}\mu
_{;\beta}e^\beta_{(1)}$.

\item (B15)$\rightarrow -\frac{1}{3}X_{II,\beta}e^\beta_{(2)}-X_{III,\beta }e^\beta_{(1)}-\frac{1}{3}(X_I+2X_{II}) (e^\beta_{(2);\beta}-a_\beta e^\beta_{(2)})-\frac{1}{3}(2X_I+X_{II})e^{(1)}_{\mu;\beta}e^\mu_{(2)} e^\beta_{(1)}
-X_{III}(e^{(1)}_{\mu;\beta}e^\mu_{(2)} e^\beta_{(2)}+e^\beta_{(1);\beta}-a_\beta e^\beta_{(1)})
=\frac{8\pi}{3}\mu _{;\beta}e^\beta_{(2)}$.

\item (B17)$\rightarrow -2 a_{(2)} {\mathcal{E}}_{(1)(1)}+2a_{(1)}{\mathcal{E}}_{(2)(1)}-E^\delta _{2;\delta}
e^2_{(2)}-\frac{Y_{I,\theta}}{3Br}+\frac{Y_{III}^\prime}{B} 
- [\frac{1}{3}(2Y_I+Y_{II})e^{(1)}_{\beta;\delta}+\frac{1}{3}(2Y_{II}+Y_I)e^\nu_{(1)} e^{(2)}_{\nu;\delta}e^{(2)}_{(\beta)}+Y_{III}(e^{(2)}_{\nu;\delta} e^\nu_{(1)}  e^{(1)}_\beta+e^{(2)}_{\beta;\delta}) ]\epsilon ^{\gamma\delta \beta} e^{(3)}_{\gamma} =-\frac{4\pi}{3}\mu
_{,\theta} e^{2}_{(2)}$.

\item (B18)$\rightarrow 2a_{(1)}{\mathcal{E}} _{(2)(2)}-2a_{(2)}{\mathcal{E}}_{(2)(1)}+E^{\delta}_{\beta;\delta}e^\beta_{(1)}
+\frac{Y_{II}^\prime}{3B}-\frac{Y_{III,\theta}}{Br}
- [-\frac{1}{3}(2Y_I+Y_{II})e^{(2)}_{\nu;\delta}e^\nu_{(1)} e^\beta_{(1)}+\frac{1}{3}(2Y_{II}+Y_I)e^{(2)}_{\beta;\delta}+Y_{III}(e^{(1)}_{\beta;\delta}- e^\nu_{(1)}  e^{(2)}_{\beta}
e^{(2)}_{\nu;\delta}) ]\epsilon ^{\gamma\delta \beta} e^{(3)}_{\gamma}=\frac{4\pi}{3}\mu_{,\beta}e^\beta_{(1)}$.
\end{itemize}

Equations (B13, B16) in \cite{1} are trivial identities, whereas (B10--B12) also become identities, due to (\ref{xs}). Finally, a detailed inspection of B1--B4, B14, B15, B17, B18, shows that these equations are identically satisfied.

\section{Conclusions}
The emission of gravitational radiation from axially symmetric sources in the context of general relativity has been described in detail in \cite{HDOC}, \cite{1}. Using the above formalism, we have shown that the transition from a state of radiation (gravitational) to an equilibrium state, is not forbidden, after a small time interval of the order of magnitude of the hydrostatic time, the relaxation time, and the thermal adjustment time.
This result is in contradiction with previous results \cite{7, DB, KN, BoNP, NPn, CTJN, Bo, BS}, that suggest that such a transition is forbidden.

The reason for such a discrepancy becomes intelligible, if we recall that in the above references, the physical properties of the source were not explicitly taken into account. Indeed, it is reasonable  to expect that any transient effect within the source, will dissapear in the time scale indicated above, due to the presence of frictional forces. Of course such phenomena are not included in any analysis of the field outside the source. Our result strengths further the relevance of the physical properties of the source, in any discussion about  the physical properties of the field. Also, it emphasizes the need to resort to global solutions, whenever important aspects about the behaviour of the gravitational field are discussed. In other words, the coupling between the source and the external field may introduce important constraints on the physical behaviour of the system, implying that  details of the source fluid cannot be left out, because they may be relevant to distant GW scattering.

The obtained result might have  been guessed from the analysis of the spacetime outside the source carried out recently in \cite{HDO}. Indeed as the main result in that reference, it is found that the absence of
vorticity implies that the exterior space-time is either static or spherically symmetric (Vaidya). Since we are evaluating the system after the thermal equilibrium has been attained, we are left with the static situation.

In spite of the above arguments, it should be clear that  the apparent  decay of the wave tail, within the time scale considered here,  must be confirmed (or denied)  by the experiment. Unfortunately though, the corresponding effects  are at least 5-6 orders of magnitude below the current detectability of  existing gyroscopes (I greatly appreciate private communication by A. Di Virgilio and Wei-Tou Ni on this issue) \cite{ginger}. Perhaps, for the GW tails, it would be much easier to be detected by the current groundbased interferometers (maybe in the third generation detectors) .

It should be kept in mind that we are dealing here, exclusively,  with the gravitational radiation produced by a source represented by a fluid distribution, due to changes in its multipolar moments. The gravitational radiation (of the ``synchrotron'' type) produced by accelerated massive particles, or the two body problem, do not belong to the  class of sources considered here.

\section{Acknowledgments}
 L.H and J.O. acknowledge financial support from the Spanish Ministry of Science and Innovation (grant FIS2009-07238) and  Fondo Europeo de Desarrollo Regional (FEDER) (grant FIS2015-65140-P) (MINECO/FEDER).

\end{document}